# Synthetic Controls with Spillover Effects: A Comparative Study[1]

*Andrii Melnychuk[2]*
*May 2024*

## Abstract

Iterative Synthetic Control Method is introduced in this study, a modification of the Synthetic Control Method (SCM) designed to improve its predictive performance by utilizing control units affected by the treatment in question. This method is then compared to other SCM modifications: SCM without any modifications, SCM after removing all spillover-affected units, Inclusive SCM, and the SP SCM model. For the comparison, Monte Carlo simulations are utilized, generating artificial datasets with known counterfactuals and comparing the predictive performance of the methods. Generally, the Inclusive SCM performed best in all settings and is relatively simple to implement. The Iterative SCM, introduced in this paper, was in close seconds, with a small difference in performance and a simpler implementation.[3]

---

[1] I am immensely grateful to my advisor, Prof. Alexis Diamond, for his guidance and encouragement throughout this project. I am also deeply thankful to Jianfei Cao, Roberta Di Stefano, Connor Dowd, and Giovanni Mellace for their valuable comments.
[2] Minerva University, melnychuk@uni.minerva.edu.
[3] All the code is available at https://github.com/Melnychuk-Andrii/Spillover-SCM.

# Table of Contents

# 1. Introduction

The Synthetic Control Method (SCM) has been widely explored and utilized in many different contexts since its first introduction by Abadie and Gardeazabal (2003). It has been further developed with methods of result evaluation, including placebos in time/space, p-value estimation, and robustness tests (Abadie et al. (2015), Abadie et al. (2010)). When measuring impact in a non-experimental setting with small pools of available units, the synthetic control method is among the best tools available. The method relies on finding a linear combination of control units that would represent the treated unit as closely as possible in the pre-treatment period. This linear combination, also known as the synthetic unit, is then used in the post-treatment period to make predictions about the counterfactual outcome for our treated unit if the treatment never happened. The pre-treatment fit is further improved by using covariates to find a pre-treatment linear combination that most resembles the treated unit on several dimensions. The method has since been applied to many empirical studies (the original paper from 2003 has been cited more than 5,900 times[4]). There has also been a fair share of modifications and improvements to the classic SCM suggested over the years (see Abadie, A. (2021) for a review).

This method relies on having a donor pool of units that were not affected by the treatment. In other words, the Stable Unit Treatment Value Assumption (SUTVA) must hold meaning that a treatment affecting one unit cannot affect any other units. This sometimes does not hold with spillover effects being common in many empirical examples. Including such units could cause significant bias in the treatment effect estimates. Traditionally, this has been addressed by removing units with (often suspected) spillover effects from the donor pool (Cavallo et al. (2013), Firpo and Possebom (2018), Kreif et al. (2016), Robbins et al. (2017), and Xu (2017)). This has its downsides though, mainly because SCM is commonly used in settings where the donor pool is small to begin with, and further reducing its size may reduce the accuracy of the method.

More recently, alternative methods have been suggested, namely, the "Inclusive SCM" by Di Stefano, R., & Mellace, G. (2024) and the "SP SCM" proposed by Cao, J. and Dowd, C. (2019). These two methods aim to make use of the donors affected by the treatment by removing the spillover-related bias from SCM estimates.

The Inclusive SCM leverages pre-intervention data to estimate the weights for constructing the synthetic control and applies a system of equations to solve for the true effects of the intervention, adjusting for the indirect impacts on the included units. This approach does not require any modification of the synthetic control estimator and can be applied using existing SC-type estimators, making it a versatile tool for researchers dealing with complex intervention scenarios.

The SP SCM extends the traditional synthetic control approach by incorporating the knowledge about potential spillover effects. This is done through an estimation process that adjusts for these indirect effects, providing a more accurate assessment of the intervention's impact. The method also includes a procedure for testing hypotheses about the presence of treatment and spillover effects, allowing for a deeper understanding of the intervention's full impact across all units involved. One limitation of this method is that it assumes a linear structure for the spillover effects. No other method explored in this paper imposes such a limitation.

---

[4] Based on information from Google Scholar (April, 2024).

Firstly, this paper introduces another method for dealing with spillover effects in SCM called "Iterative SCM". This modification involves using the Synthetic Control Method to "clean up" donors affected by spillover effects and then using the synthetic versions of these donors in the donor pool for a final SCM run with our treated unit in question. This modification also allows for any SC-type estimators to be used instead of the classic SCM with no limitations.

Secondly, this paper compares the SCM modifications designed to deal with spillover control units. Five methods have been compared here: Unrestricted SCM, Restricted SCM, Inclusive SCM, SP SCM, and Iterative SCM.

- Unrestricted SCM is defined here as running the synthetic control without any modifications to the method and without excluding potentially affected control units from the donor pool. It's worth noting that knowingly leaving potentially affected control units is not an appropriate use of the SCM since the method was designed to rely on SUTVA and the presence of spillover effects violates this assumption.
- Restricted SCM is defined here as running the synthetic control without any modifications to the method and excluding all potentially affected units from the donor pool. This is also the suggested method in Abadie et al. (2010) and Grossi et al. (2020) and it is most commonly used in literature for dealing with spillover effects when utilizing SCM.
- Inclusive SCM is the SCM modification described in Di Stefano, R., & Mellace, G. (2024).
- SP SCM is the SCM modification described in Cao, J. and Dowd, C. (2019).
- Iterative SCM is the modification introduced in this paper and further discussed in Section 2.

The rest of the paper is organized as follows: Section 2 describes the Iterative SCM; Section 3 contains the methodology, including artificial data simulation implemented to test the methods; Section 4 has Monte Carlo simulation results; Section 5 describes empirical results; Section 6 is for limitations; Section 7 contains conclusions.

Table 1. Key differences between SCM modifications for spillover cases.

| Method | Uses spillover-affected units for prediction | Assumes a specific spillover structure | Requires access to donor weights of synthetic units |
|---|---|---|---|
| Unrestricted SCM | Yes* | No | No |
| Restricted SCM | No | No | No |
| SP model | Yes | Yes** | Yes |
| Inclusive SCM | Yes | No | Yes |
| Iterative SCM | Yes | No | No |

\* Unrestricted SCM includes spillover-affected units in the donor pool but does not control for the spillover effects in any way. This results in biased estimates.

\*\* SP model assumes a linear structure for spillover effects.

## 2. Iterative Synthetic Control Method

A novel component of this study is the introduction of "Iterative SCM". This modification involves a fairly straightforward way to remove spillover effects from the donor pool. It's done by creating a synthetic control for each potentially affected donor and then imputing the real potentially affected donors with their synthetic counterparts in the donor pool of the final SCM run.

It is worth noting that when proposing this method, I did not expect outstanding performance out of it. More than anything, I wanted to see how close this simple approach would come to the results achieved by some of the more involved and thought-through methods (Inclusive SCM, SP SCM).

Two variations of the "Iterative SCM" have been explored during the development of this method.

The first is whether to impute pre-treatment donor outcome values with their synthetic counterparts, further referred to as "Replace pre-treatment data". If yes, then pre-treatment outcome data will be imputed with its synthetic counterpart when "cleaning" spillover units. If no, then pre-treatment data of spillover units will be preserved and only post-treatment data will be imputed.

The second is whether to use already "cleaned" donors for subsequent donor cleanups, further referred to as "Use already cleaned controls". If yes, then the first donor "cleaning" SCM uses only known "good" donors, subsequent donor "cleanings" use known "good" donors as well as already "cleaned" donors. If no, then each donor "cleaning" SCM uses only known "good" donors, which is the same as Restricted SCM. Since each variation can either be "on" or "off", this gives a total of 4 possible combinations of these variations.

## 3. Methodology

### 3.1. Monte Carlo simulations

Monte Carlo simulations were utilized to generate artificial datasets with known counterfactuals. 5 methods have been executed on each artificial dataset: Unrestricted SCM, Restricted SCM, Inclusive SCM, SP SCM, and Iterative SCM (see the end of Section 1 for descriptions of methods). The configuration list for data generation consisted of all possible combinations of the parameter values provided in Table 1 (384 combinations in total). For every data generation configuration in the list, 100 datasets were generated and used as input for all the methods tested. Prediction errors (PE) were calculated for each method on each configuration. PEs were defined as the difference between the real $y_0$ counterfactual and the predicted $y_0$ for the treated unit (y0 is the outcome without treatment for a given unit; y1 is the outcome with treatment for a given unit). The resulting post-treatment prediction errors were aggregated together into vectors. A vector was allocated for every method-configuration combination (resulting in 384*5 = 1,920 PE vectors, each with 100 values).

Table 2. Parameter values for data generation.

| Parameter | Values |
| --- | --- |
| Pre-treatment period length | 10; 15 |
| Total number of control units | 5; 9 |
| Ratio of spillover units out of all control units | 0.11; 0.33; 0.67; 0.9 |
| Treatment effect | 1.8; 3; 5 |
| Ratio of spillover effect to treatment effect | 0.1; 0.3; 0.6; 0.9 |
| Data Generation Process (DGP) cases | Stationary; I(1) |

The post-treatment period was fixed as 1 timepoint for all simulated datasets. After collecting all the post-treatment gaps, the mean squared prediction errors (later referred to as MSPE) were calculated for each unique method & configuration combination using the PE vectors constructed earlier. This resulted in 1,920 unique MSPE data points. Each MSPE data point is the mean of 100 prediction gaps squared (100 datasets for every given data generation parameter combination, each with 1 post-treatment prediction).

### 3.2. Data generation processes (DGPs)

The underlying data generation processes for the Stationary case and I(1) case were replicated based on Cao, J. and Dowd, C. (2019) utilizing a factor model structure for both. The only difference made to the models in this paper is that the treatment and spillover effects are not fixed at 5 and 3 respectively but instead defined by the data generation parameters as described earlier.

## 4. Simulation Results

### 4.1. Iterative SCM variations

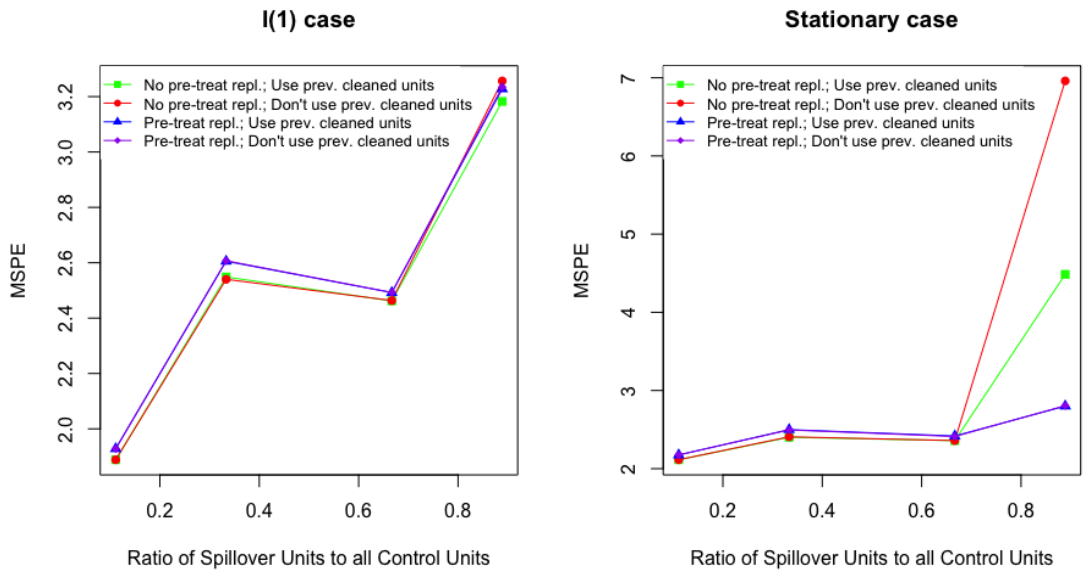

Figure 1. Post-treatment MSPE of Iterative method variations depending on the ratio of spillover units to total control units.

Table 3. Post-treatment MSPE of Iterative method variations. Mean of all data points. Stationary case.

| Stationary case | | Use already cleaned controls | |
|---|---|---|---|
| | | Yes | No |
| Replace pre-treatment data | Yes | 2,4735 | 2,4722 |
| | No | 2,8401 | 3,4604 |

Table 4. Post-treatment MSPE of Iterative method variations. Mean of all data points. I(1) case.

| I(1) case | | Use already cleaned controls | |
|---|---|---|---|
| | | Yes | No |
| Replace pre-treatment data | Yes | 2,5634 | 2,5656 |
| | No | 2,5204 | 2,5371 |

All the variations performed very similarly in most cases, with the only notable exception being a high spillover-to-control ratio in the Stationary DGP case. Based on this, the "Replace pre-treatment data; Use already cleaned controls." variation was chosen for further analysis as a variation with consistently high performance.

## 4.2. Total number of Control Units in the dataset

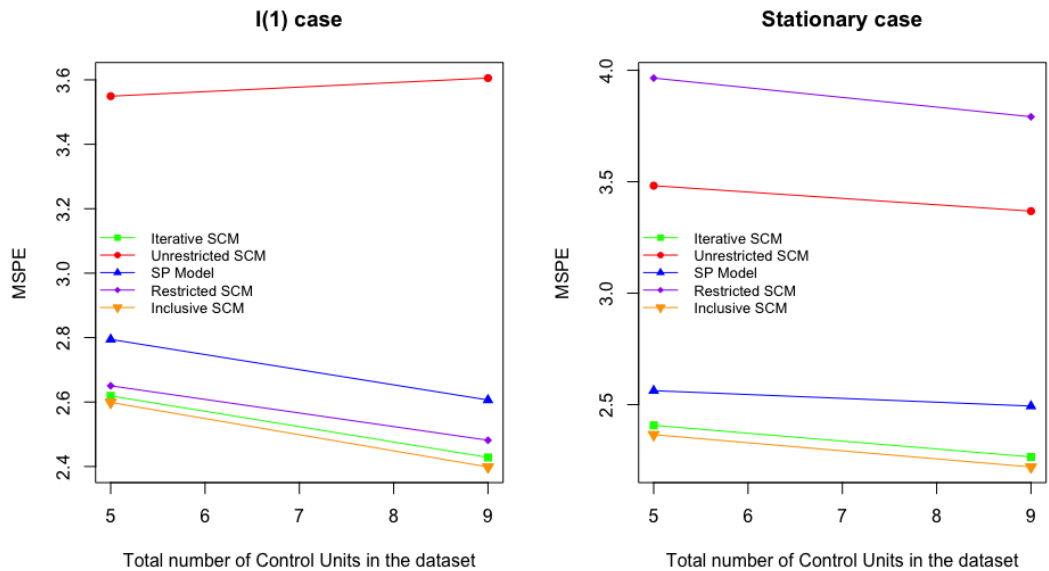

Figure 2. MSPE results for all methods depending on the total number of control units in the dataset.

Based on the Stationary case, we can observe a common trend among all methods, the more units there are the better the performance. That being said, it's interesting to note that the performance of the Restricted method seems to benefit a little more from a larger control unit pool. That is most likely due to the nature of the method (removing all spillover units and just running SCM on what's left). The larger the donor pool, the more controls will be left over for the Restricted method to work with. Based on the I(1) case, a similar picture arises with the only exception being Unrestricted SCM that degrades slightly in performance.

## 4.3. Ratio of Spillover Units to all Control Units

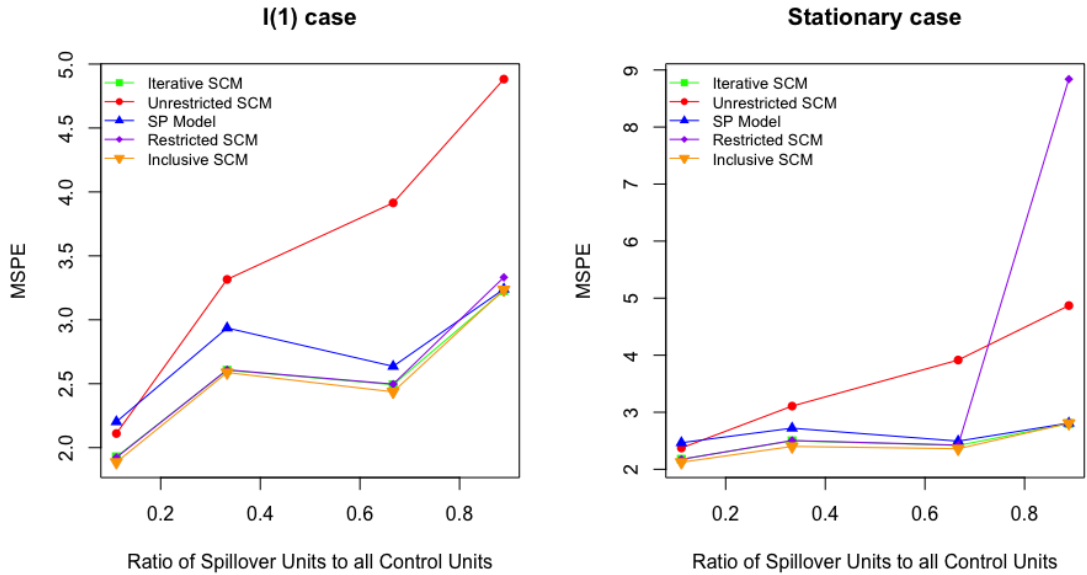

Figure 3. MSPE results for all methods depending on the ratio of spillover units to the total number of controls in the dataset.

From both DGP cases, as the ratio of spillover units increases, the bias of the Unrestricted SCM also increases. The more spillover units there are, the more bias gets picked up by the Unrestricted SCM. Additionally, SP SCM matches the performance of the Inclusive and Iterative methods when the ratio is high. When the ratio is <= ~0.6, Restrictive SCM seems to do as good a job as the SP SCM, Inclusive, or Iterative. For the Stationary case, an interesting trend emerged where the performance of the Restricted method was remarkably close to the Inclusive and Iterative methods up until a certain "break-off" point. When the ratio of spillover units to the total number of controls was 0.9, the performance of the Restricted method degraded significantly. Additionally, the performance of all 3 SCM modifications designed to deal with spillover effects (Iterative, Inclusive, SP SCM) seems to converge to the same value when the ratio is high.

## 4.4. Treatment Effect size

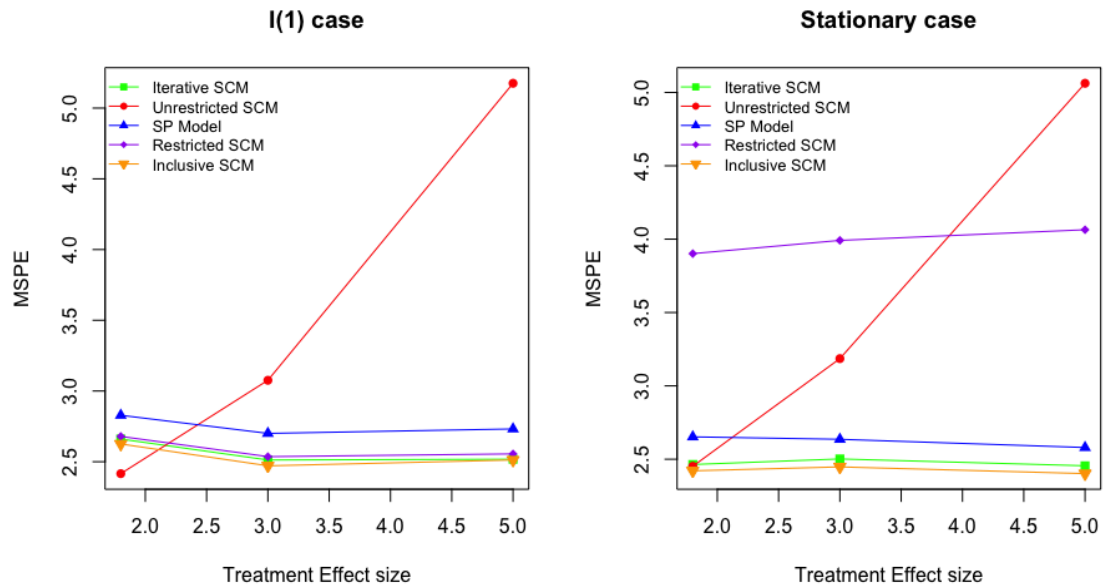

Figure 4. MSPE results for all methods depending on the size of the treatment effect in the dataset.

As expected, for both DGP cases the greater the treatment effect, the more bias is picked up by the Unrestricted SCM because the spillover effect size is defined as a fraction of the treatment effect, so the greater the treatment effect, the greater the spillover effects. Also, the Restricted method seems to perform poorly on the Stationary case when compared to the I(1) case.

## 4.5. Ratio of Spillover Effect to Treatment Effect

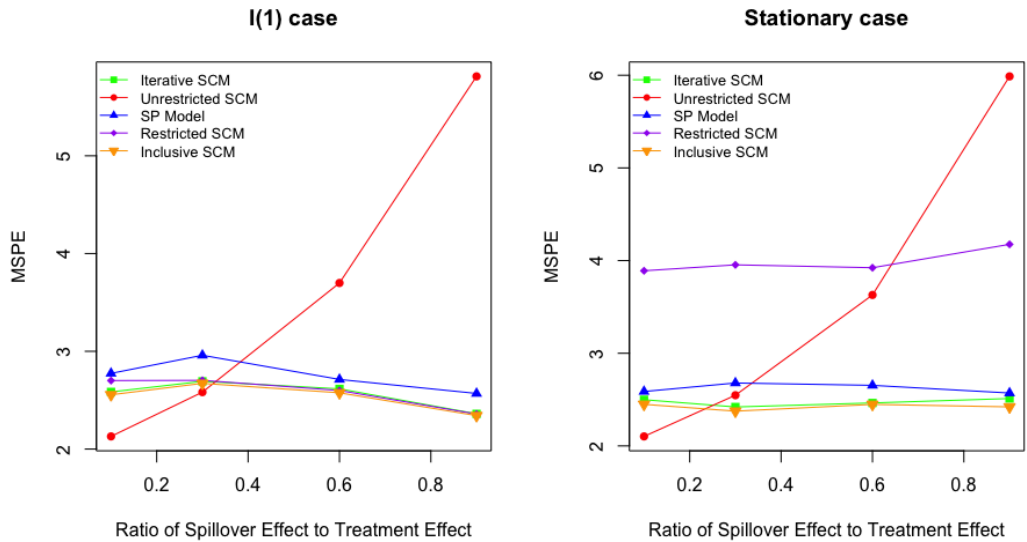

Figure 5. MSPE results for all methods depending on the ratio of the spillover effect to the treatment effect in the dataset.

A similar picture emerges to the previous plot, where the greater the spillover effect the more bias is picked up by the Unrestricted SCM. Also, similarly to the previous plot, the Restricted method seems to perform poorly on the Stationary case when compared to the I(1) case.

## 4.6. Pre-treatment period length

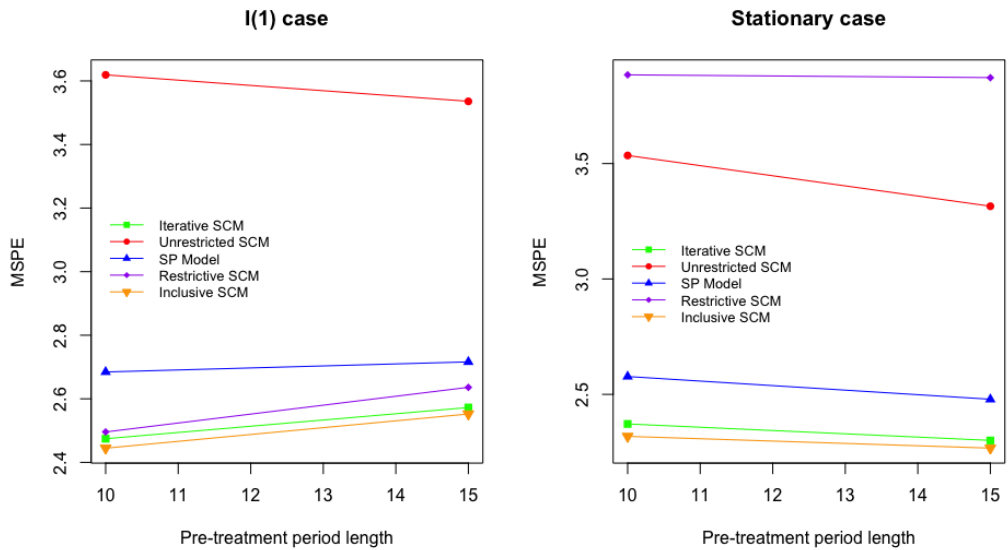

Figure 6. MSPE results for all methods depending on the length of the pre-treatment period in the dataset.

A notable observation for the I(1) case is that the Iterative, Inclusive, Restrictive, and SP SCM all degrade in performance slightly as the pre-treatment period increases. The Unrestricted SCM improves slightly in both DGP cases. In the Stationary DGP case, all methods except for the Restrictive SCM perform slightly better with increased pre-treatment period length.

## 4.7. Comparison of DGP cases

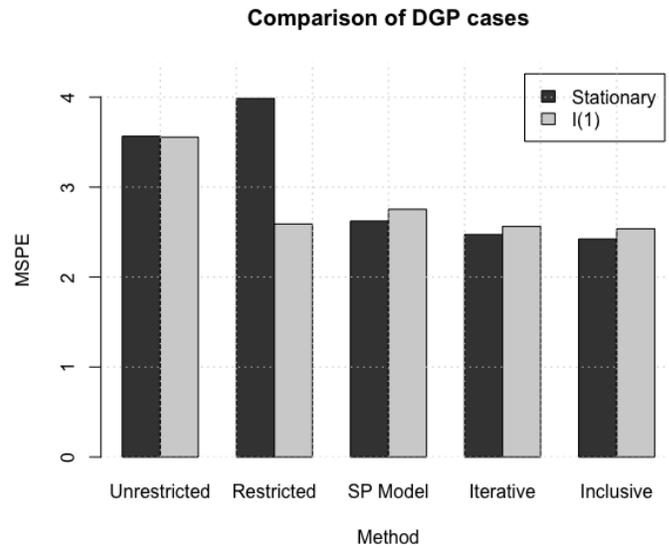

Figure 7. MSPE results for all methods depending on the DGP in the dataset.

Table 5. MSPE results for all methods depending on the DGP in the dataset.

| Method | MSPE for Stationary | MSPE for I(1) |
|---|---|---|
| Unrestricted SCM | 3.5666 | 3.5557 |
| Restricted SCM | 3.9856 | 2.5899 |
| SP model | 2.6225 | 2.7534 |
| Iterative SCM | 2.4735 | 2.5634 |
| Inclusive SCM | 2.4233 | 2.5362 |

The stationary case seems to provide a greater challenge for the Restricted method, while in the I(1) case the Restricted method performs similarly to all 3 SCM modifications designed to deal with spillover effects (Iterative, Inclusive, SP SCM).

# 5. Empirical Examples

## 5.1. California's Tobacco Control Program

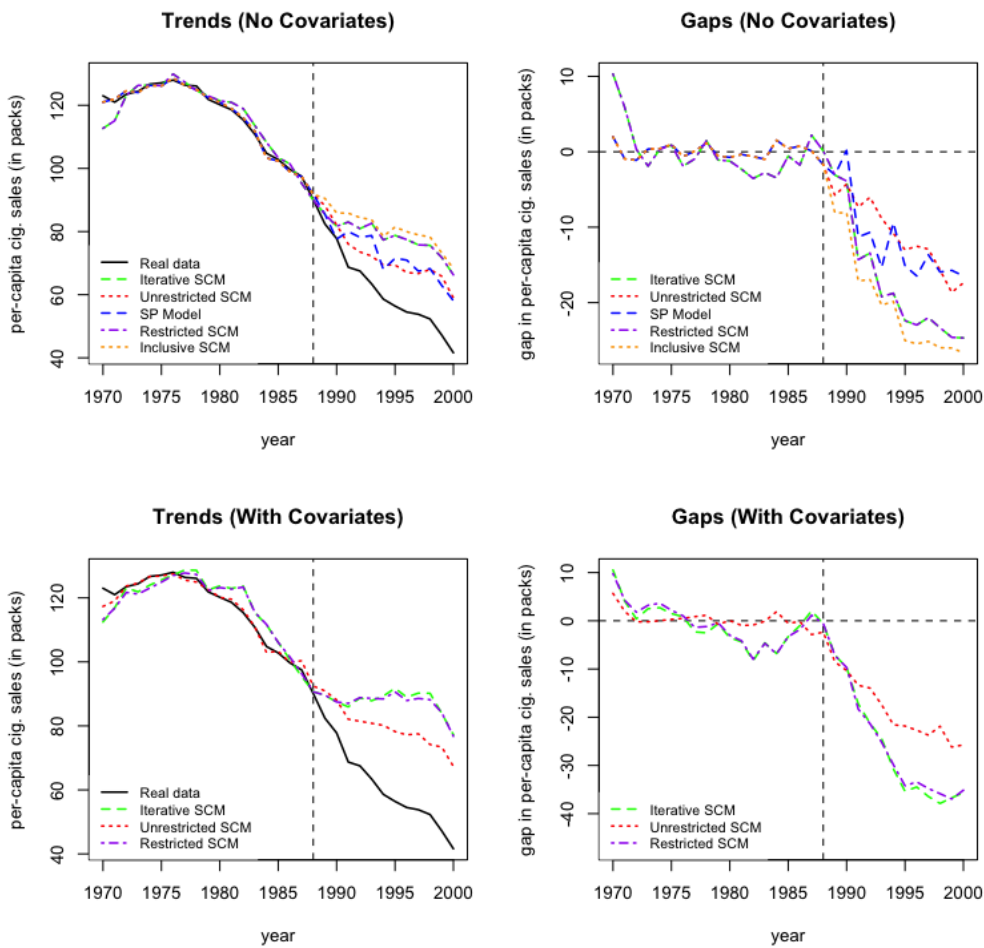

Figure 8. Estimates for the California dataset[5]. The vertical line at year 1988 represents the treatment year.

The California Tobacco Control Program initiated through Proposition 99 in 1988, increased the cigarette excise tax by 25 cents per pack, directing the revenue towards health and anti-smoking education. It also supported media campaigns and spurred numerous local non-smoking ordinances. This dataset was first explored using SCM by Abadie et al. (2010) and further used as an example by Cao, J. and Dowd, C. (2019). The setup used in this study follows the general SCM settings by Abadie et al. (2010) and spillover unit assignment by Cao, J. and Dowd, C. (2019).

---

[5] Inclusive SCM and SP SCM are missing on all "With Covariates" plots in empirical examples. Some difficulties have been encountered while implementing these two methods with covariates. Communication with both methods authors is ongoing to finalize the implementation.

The root mean squared difference between the Unrestricted and Iterative SCM is 9.42 cig. pack sales per capita. So, based on the Iterative SCM, the mean treatment effect of the program between 1989 and 2000 is a reduction in cig. sales of 28.46 packs per capita (37% reduction) as opposed to the estimate of 19.03 packs per capita prediction (25% reduction) made by Abadie et al. (2010).

## 5.2. West Germany Reunification

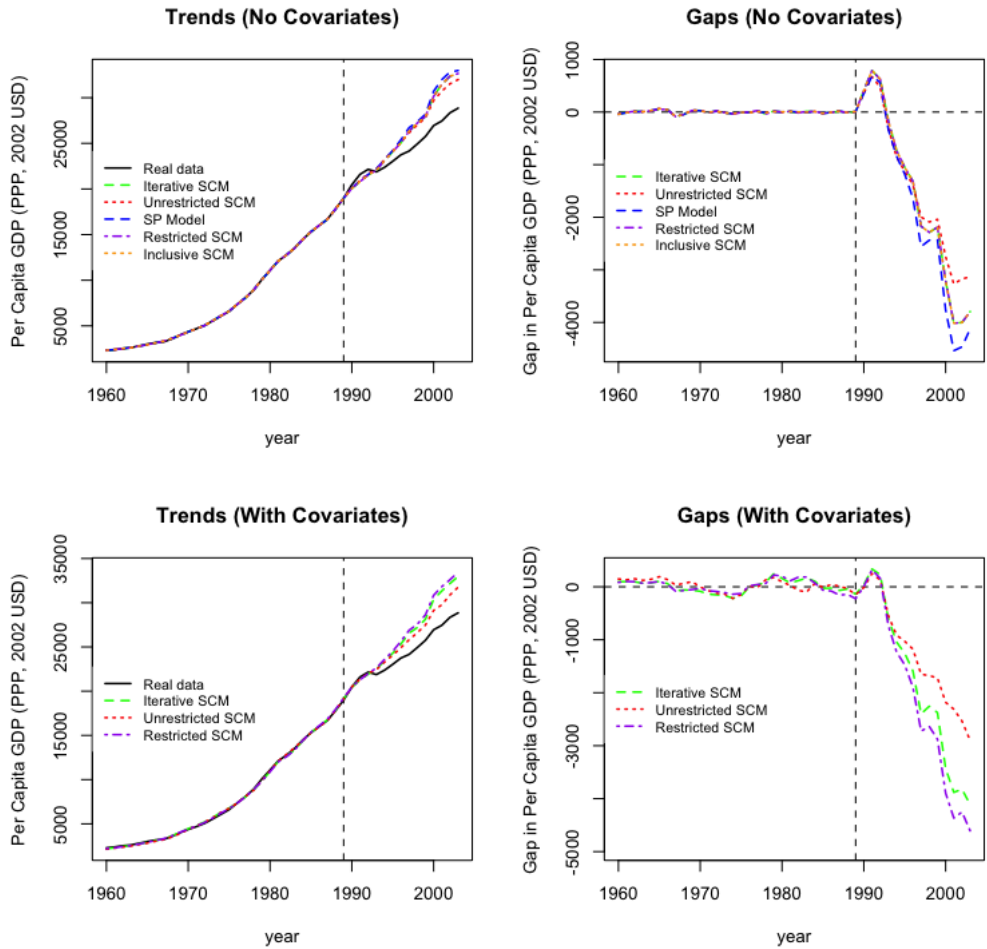

Figure 9. Estimates for the West Germany dataset. The vertical line at the year 1989 represents the treatment year.

This example estimates the economic impact of the 1990 German reunification on West Germany. It was first explored using SCM by Abadie et al. (2015) and further used as an example by Di Stefano, R., & Mellace, G. (2024). The setup used in this study follows the general SCM settings by Abadie et al. (2015) and spillover unit assignment by Di Stefano, R., & Mellace, G. (2024).

The root mean squared difference between the Unrestricted and Iterative SCM is 371 USD (2002) in GDP per capita. So, based on the Iterative SCM, the mean treatment effect of the reunification between 1990 and 2003 is a reduction in GDP per capita of ~1,970 USD (9.8% reduction) as opposed to the estimate of ~1,600 USD prediction (8% reduction) made by Abadie et al. (2015).

## 5.3. Basque Country Terrorism

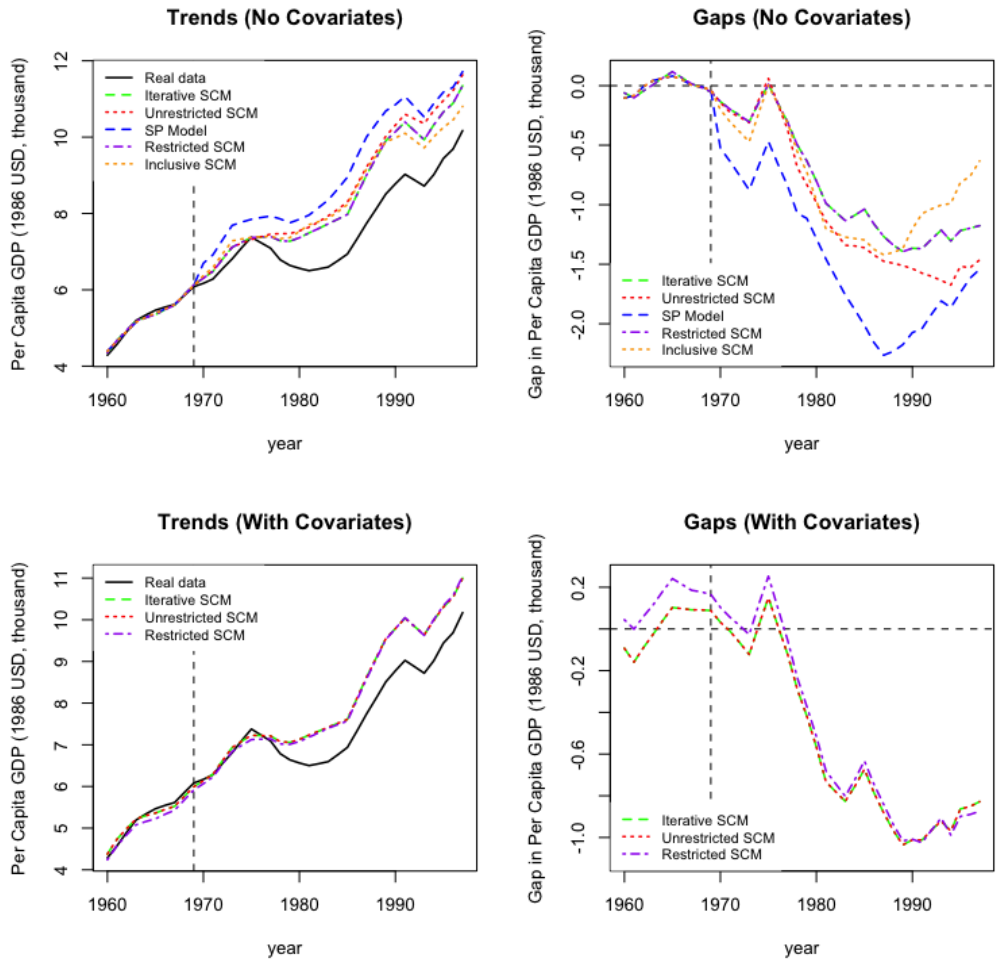

Figure 10. Estimates for the Basque dataset. The vertical line at year 1969 represents the treatment year.

This example uses the Basque Country as a case study for the effect of a terrorism outbreak in the late 1960s on the economy of the region. First explored using SCM by Abadie and Gardeazabal (2003), this example is the first and most cited application of the Synthetic Control Method. All the regions sharing a land border with Basque were considered spillover units (Navarra, La Rioja, Castile and León, and Cantabria). There is no evidence of spillover effects in this example. It is worth noting that without covariates all the methods tested vary widely in their post-treatment predictions when compared to the "with covariates" case.

## 5.4. 2014 Eastern Ukrainian Crisis

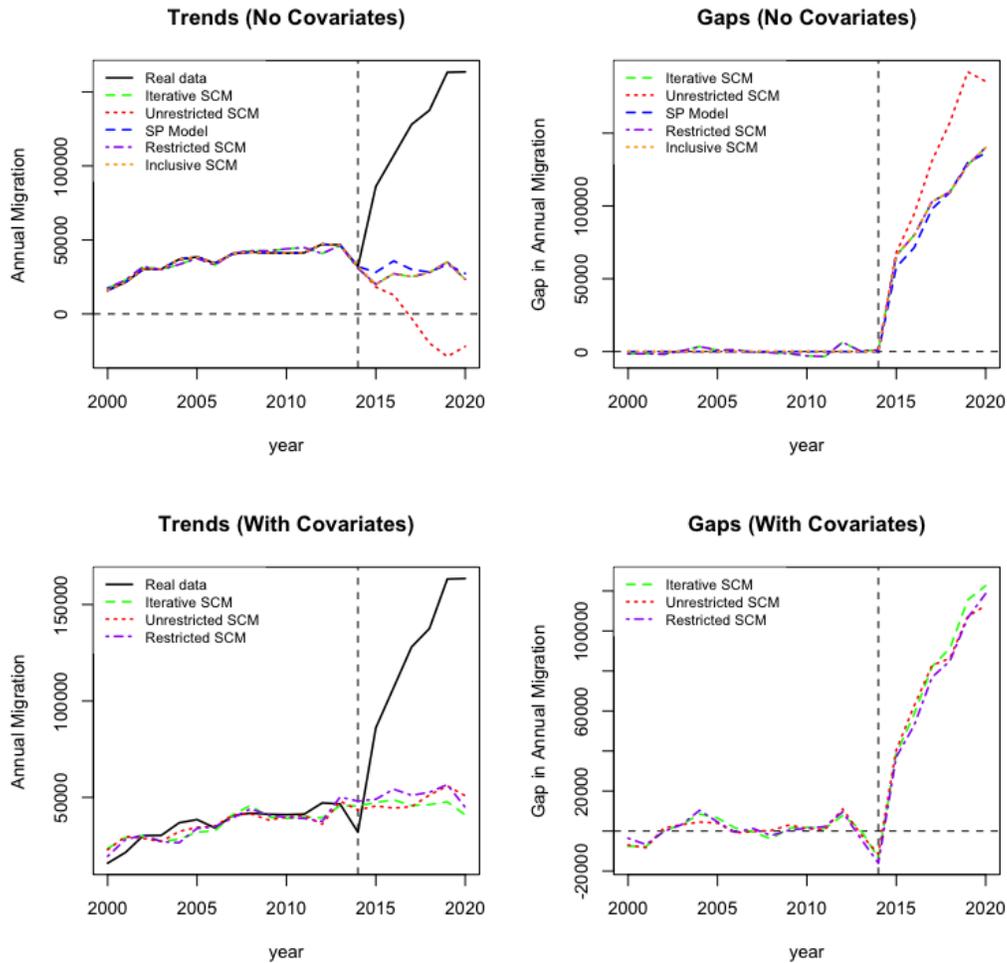

Figure 11. Estimates for the Ukraine Crisis dataset. The vertical line at year 2014 represents the treatment year.

This example uses migration data in Europe to study the effects of the 2014 Eastern Ukrainian crisis on immigration rates in Poland, explored using SCM by Besterekova (2024). The most notable observation here is the poor post-treatment performance of the Unrestricted SCM without covariates. This shows how important covariates can be when using SCM on empirical data since in both cases the pre-treatment fit is reasonable but without covariates, the post-treatment predictions are quite inaccurate for the Unrestricted SCM. It's also worth noting that the other SCM modifications tested did not degrade in predictive performance as much when excluding covariates in this dataset.

# 6. Limitations

One limitation of this study is that covariates were not utilized in the artificial datasets. Covariates are commonly used in empirical research to improve the pre-treatment fit of the synthetic unit. It would be interesting to see whether any of the methods benefit more or less than the others from having covariates.

Another limitation is related to the data generation processes used. Based on the results in Section 4.7, it is easy to see that for some datasets (i.e. I(1) case), SCM modifications aimed to deal with spillover effects provide little to no improvement when compared to just removing all spillover units (Restricted SCM). That being said, for some datasets, the improvement is very significant (i.e. Stationary case). While the data generation processes for real-world data are seldom known, it is nonetheless an important limitation because only 2 different data generation processes were tested in this study. Additionally, in this study, the spillover effect size was the same for all the spillover units. This would be very rare with real-world data. Different data generation processes should be explored to make a stronger conclusion about the performance of the methods.

Also, cases of under/over-specification of spillover units have not been explored. However, this is a fairly significant case for future empirical research since the exact number of and assignment of spillover effects is usually unknown. After studying this, practical guidelines could be made for when to over-specify or under-specify spillover units.

Placebos in time and space, as well as Robustness checks, have not been explored either. These are common methods for validating SCM results in empirical research. Placebos in space, introduced by Abadie and Gardeazabal (2003), consist of reassigning the treatment flag to units that have never been treated to evaluate whether the SCM would predict a treatment effect when there should be none. Placebos in time, introduced by Abadie et al. (2010), rely on reassigning the time of treatment to one before the actual treatment to evaluate whether a gap in the outcome would be predicted for times when treatment did not happen. Robustness checks, introduced by Abadie et al. (2015), rely on removing control units one by one from the donor pool to see how the SCM prediction changes without a particular unit and evaluate to what extent any particular control unit drives the results. It would be interesting to see whether these validation methods reveal any additional differences between the methods explored in this study.

Lastly, 4 limitations have been intentionally imposed due to the computational complexity of running thousands of Synthetic Controls. Firstly, only pre-treatment period lengths of 10 and 15 were explored. A greater range could reveal more about the performance of the methods. For one, it would be interesting to see whether the performance improvements of the Restricted SCM with larger pre-treatment periods would allow it to catch up with other methods or whether it would plateau (see Section 4.6). Secondly, only total control unit counts of 5 and 9 were explored. A greater range here could also reveal more about the performance of the methods. Thirdly, the "no spillover units" case was not included. It doesn't make much sense to run methods designed to deal with spillover effects when there are none, but the primary reason why this case was not included is that 4 out of 5 methods included in this paper would have identical performance since they would all end up executing the same classic SCM configuration behind the scenes due to the way they are defined. The only method with slightly different results would be the SP SCM, however, its performance was compared to the Unrestricted SCM in a "no spillover units" case by Cao, J. and Dowd, C. (2019), and the SP SCM had similar mean prediction errors to the Unrestricted SCM with slightly higher prediction error variance in all data generation configurations tested. Finally, only 100 iterations per data generation configuration were utilized. Using a higher number would increase the reliability of the results.

## 7. Conclusions

In this study, a new modification of the SCM called "Iterative SCM" was explored. A variation of the Iterative method in which we replace the pre-treatment data and use previously "cleaned" donors to "clean" subsequent donors had the best performance out of all 4 variations tested. This variation, later referred to as just "Iterative SCM", was further compared in post-treatment prediction performance to 4 other methods (Unrestricted SCM, Restricted SCM, Inclusive SCM, SP SCM) using both artificially generated datasets with known actual treatment effects and empirical data with unknown actual treatment effects.

The Inclusive SCM consistently comes out on top with the Iterative SCM as a close second based on MSPEs. The SP SCM seems to match the performance of the Inclusive and Iterative SCMs when the ratio of spillover units to the total number of units is high. That being said, the difference in performance for smaller ratios of spillover units is not particularly great, indicating that there is likely little practical difference between the 3 methods in terms of prediction accuracy. Additionally, when the ratio of spillover units is <= ~0.6, there is little to no benefit in using SCM modifications designed to deal with spillover effects in the first place because the Restricted SCM has very similar performance while offering a significantly simpler methodology. Further research would be needed to make a more robust conclusion about any differences in the performance of these methods. One possible explanation could be that the Inclusive SCM has more precision (and thus a smaller MSPE) while SP SCM is more robust to spillover unit misspecification.

If there are fewer spillover units than clean control units, then the Restricted SCM (removing all spillover units from the donor pool) appears to perform just as well as the Iterative SCM, Inclusive SCM, or the SP SCM. So the SP SCM, Inclusive, and Iterative SCM are only really needed in cases where the spillover effects are prevalent (with more than ~60% of controls being affected). In such a case, any one of the 3 spillover-focused methods (SP SCM, Inclusive, and Iterative SCM) can be used. Iterative SCM has almost identical predictive performance to Inclusive SCM but is noticeably simpler in implementation. One benefit that Iterative SCM has over the Inclusive SCM and SP SCM is that any estimator can be used in place of the classic SCM. For Inclusive SCM to work, SCM weights need to be accessed, but it's not necessary for the Iterative approach making it slightly more flexible.

Practically speaking, a procedure adopted from the one described by Di Stefano, R., and Mellace, G. (2024) is presented in Figure 12 and recommended for future empirical studies dealing with potential spillover effects. It's worth noting that these practical recommendations are based on the literature cited in this paper together with all the findings from this paper. See Section 6 (Limitations) before running analyses based on these findings.

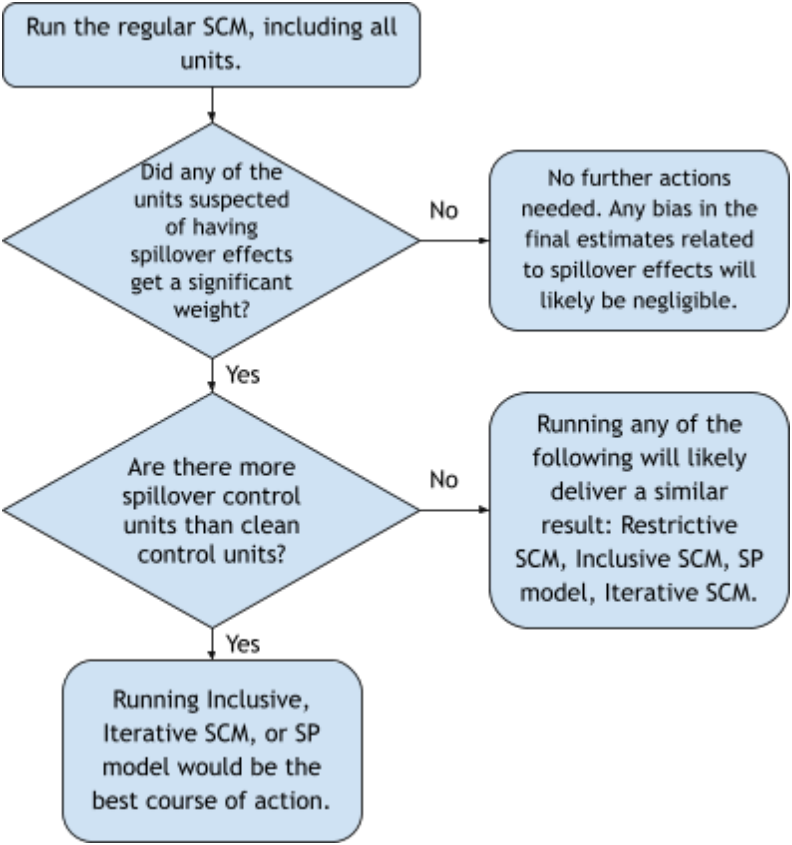

Figure 12. Algorithm for deciding when to use SCM modifications made to deal with spillover effects.